\documentclass[
]{ceurart}

\usepackage{listings}
\newcommand{\art}{{\sc ART}}
\newcommand{\lart}{{\sc L'ART}}

\begin{document}

\copyrightyear{2023}
\copyrightclause{Copyright for this paper by its authors.
  Use permitted under Creative Commons License Attribution 4.0
  International (CC BY 4.0).}

\conference{London'23: Workshop on Logic Programming and Legal Reasoning,
  July 09--10, 2023, London, UK}

\title{How well do SOTA legal reasoning models support abductive reasoning?}

\tnotemark[1]

\author[1]{Ha Thanh Nguyen}[%
orcid=0000-0003-2794-7010,
email=nguyenhathanh@nii.ac.jp
]
\cormark[1]
\address[1]{National Institute of Informatics (NII),
  2-1-2 Hitotsubashi, Chiyoda City, Tokyo, Japan}

\author[2]{Randy Goebel}[%
orcid=0000-0002-0739-2946,
email=rgoebel@ualberta.ca
]
\address[2]{University of Alberta, 116 St \& 85 Ave, Edmonton, AB T6G 2R3, Canada}

\author[3]{Francesca Toni}[%
orcid=0000-0001-8194-1459,
email=f.toni@imperial.ac.uk,
]
\address[3]{Imperial College London, Exhibition Rd, South Kensington, London SW7 2BX, United Kingdom}

\author[4]{Kostas Stathis}[%
orcid=0000-0002-9946-4037,
email=Kostas.Stathis@rhul.ac.uk,
]
\address[4]{Royal Holloway University of London, Egham Hill, Egham TW20 0EX, United Kingdom}

\author[1]{Ken Satoh}[%
orcid=0000-0002-9309-4602,
email=ksatoh@nii.ac.jp,
]
\cortext[1]{Corresponding author.}

\begin{abstract}
  We examine how well the state-of-the-art (SOTA) models used in legal reasoning support abductive reasoning tasks. Abductive reasoning is a form of logical inference in which a hypothesis is formulated from a set of observations, and that hypothesis is used to explain the observations. The ability to formulate such hypotheses is important for lawyers and legal scholars as it helps them articulate logical arguments, interpret laws, and develop legal theories. Our motivation is to consider the belief that deep learning models, especially large language models (LLMs), will soon replace lawyers because they perform well on tasks related to legal text processing.  But to do so, we believe, requires some form of abductive hypothesis formation.  In other words, while LLMs become more popular and powerful, we want to investigate their capacity for abductive reasoning. To pursue this goal,  we start by building a logic-augmented dataset for abductive reasoning with 498,697 samples and then use it to evaluate the performance of a SOTA model in the legal field. Our experimental results show that although these models can perform well on tasks related to some aspects of legal text processing, they still fall short in supporting abductive reasoning tasks.
\end{abstract}


\maketitle

\section{Introduction}
\label{sec:intro}

The rise of transformer-based deep learning models \cite{vaswani2017attention} has brought remarkable advancements in natural language processing (NLP), including tasks related to legal text processing \cite{ijcai2020p484,chalkidis2020legal,nguyen2020jnlp,tran2020encoded,yoshioka2021bert,nguyen2022attentive}. These advancements have the potential to improve access to justice and legal services for underserved communities, and to enhance the efficiency and accuracy of judicial processes.

\begin{figure*}
\center
\includegraphics[width=.6\textwidth]{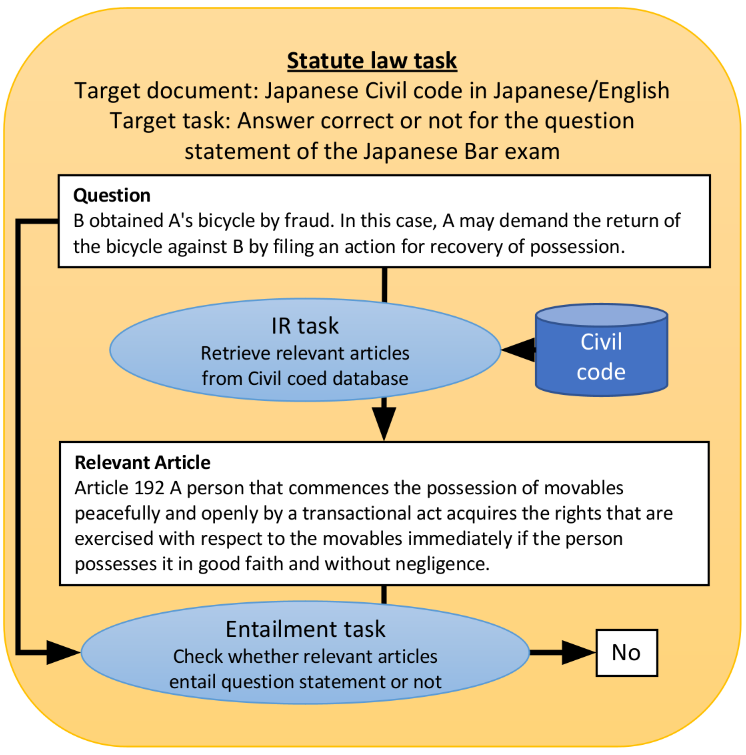}
\caption{An example of statute law retrieval task requiring abductive reasoning skill in COLIEE Competition~\cite{Rabelo_2022}. The truth value (No) of the statement is determined by assuming that the statement is either true or false, and then searching for the appropriate legal article that provides the best explanation for that assumption. In this case, the relevant article is Article 192, which provides an exception to A's ability to demand the return of the bicycle against B through an action for recovery of possession. Abductive reasoning is utilized in determining that the statement's value is "No" based on the application of Article 192 and the exception it presents.}\label{fig:example}
\end{figure*}

However, a critical component of legal intelligence is abductive reasoning, which is of paramount importance for lawyers and legal scholars in formulating logical arguments, interpreting laws, and developing legal theories \cite{abimbola2000abductive,schum2002species}. 
Figure \ref{fig:example} is an example of statute law retrieval task requiring abductive reasoning skill in the existing COLIEE Competition~\cite{Rabelo_2022}. 
Since transformer-based models have shown promising results in various NLP tasks within the legal field, it is essential to evaluate their performance on tasks involving abductive reasoning to better understand their capabilities and limitations as AI tools in the legal field.

Existing literature has explored the development and evaluation of transformer-based models for legal text processing tasks such as legal document retrieval, summarization, entailment, and question-answering. Yet, there is a gap in evaluating these models for the crucial task of abductive reasoning. To address this gap, we first articulate an empirical method to reveal the role of abductive reasoning in legal information processing. We identify an appropriate corpus of examples by examining the \art{} dataset~\cite{bhagavatula2020abductive}, which is the first large dataset for abductive reasoning tasks, and ${\alpha}NLI$ task, which investigates the viability of language-based abductive reasoning. 

In this paper, we focus on:
\begin{itemize}
    \item Enhancing the reliability of the dataset for abductive reasoning through task redefinition and data augmentation.
    \item Evaluating the performance of a state-of-the-art (SOTA) transformer-based model in the legal field on abductive reasoning tasks.
\end{itemize}

Our experimental results show that although the selected SOTA model can perform well on tasks related to legal text processing, it still falls short in supporting abductive reasoning tasks, shedding light on an important limitation of these models in legal reasoning. This study provides a more comprehensive understanding of the capabilities of transformer-based models in the legal domain, emphasizing the importance of abductive reasoning, which is often overlooked in related research.

\section{Background}
\label{sec:background}
\subsection{Legal Text Processing}
There has been significant research on the use of artificial intelligence and machine learning (AI) in the legal field in recent years. This research has resulted in the development of a number of state-of-the-art models that are able to perform various tasks related to legal text processing, such as contract risk analysis and case law retrieval.

Deep learning models have the ability of automated latent feature extraction, 
which allows us to use these models not only for similarity matching tasks but also in other semantic matching tasks such as question answering ~\cite{kien-etal-2020-answering,asai-etal-2021-xor}, machine reading comprehension \cite{nie-etal-2019-revealing,lee-yeung-2021-text}, image retrieval \cite{krojer-etal-2022-image} and entity matching \cite{lin-etal-2020-triggerner,ritter-etal-2011-named}. 
These tasks are all important for the general challenge of legal reasoning.

In the legal retrieval task, legal documents are usually structured and existing systems are designed to retrieve relevant legal texts (e.g., regulations) based on a given query.  This task is an essential component of intelligent legal counsel systems and commonly appears in legal automated processing competitions \cite{Rabelo_2020,Rabelo_2022,Thanh_2021,Nguyen_2022}. 
One of the challenges in legal information retrieval is that the available data is usually very limited.
This is why the current best systems often need to be based on some supportive rules or data augmentation methods.
As a consequence, Deep learning with transfer learning methods has been successfully applied to this problem in a number of ways \cite{kien-etal-2020-answering,ijcai2020p484,Nguyen_transformer_2022,louis-spanakis-2022-statutory,Vuong_2022}. 

Retrieval tasks are foundational to many other legal text processing tasks. For example, contract analysis often involves retrieving relevant provisions or clauses from a contract based on a given query. Similarly, case law retrieval involves identifying and retrieving relevant case law and legal precedent based on a given query. These tasks require making inferences and educated guesses about what information is likely to be relevant, based on the characteristics of the query and the available data.

In addition, many other legal text-processing tasks, such as legal document classification and summarization, are also related to retrieval. For instance, classifying a legal document as relevant or irrelevant to a given case may involve retrieving similar documents and using them as a reference. Similarly, summarizing a legal document may involve retrieving relevant information and condensing it to a shorter form.
Overall, retrieval tasks form the foundation for many legal text processing tasks, and the ability to perform retrieval effectively is essential for the development of intelligent legal counsel systems and other AI tools for the legal field.

Note that retrieval tasks and abductive reasoning are closely related. In retrieval tasks, the goal is to identify and retrieve relevant information from a database or collection of documents based on a given query. This requires making an educated guess or inference about what information is likely to be relevant, based on the characteristics of the query and the available data.
Similarly, in abductive reasoning, the goal is to construct an argument or form a hypothesis based on a set of observations and a limited amount of information. This also requires making an educated guess or inference based on the available evidence, in order to explain the observations and arrive at a plausible conclusion.

\subsection{Abductive Reasoning} 
\label{sec:ar}
Our approach to test deep learning models based on transformers \cite{vaswani2017attention} for reasoning is focused on abductive reasoning tasks, because abductive reasoning is such an important aspect of human reasoning~\cite{kakas1992abductive}.
In a typical abductive reasoning problem, one is given a set of observations, and the goal is to identify a hypothesis that can best explain the observations.
This process can be divided into two steps: 1) potential hypothesis identification, and 2) hypothesis evaluation. In the first step, we need to identify a small set of potential hypotheses that are likely to explain the observations. In the second step, we need to evaluate the potential hypotheses and rank them within the current context of use, e.g., assess a set of alternative hypotheses w.r.t. which provides the basis for a ``best'' explanation.

Because the complexity of the real world is high, it is impossible to consider all potential hypotheses. Instead, we typically assume that a set of potential hypotheses is given, and the evaluation of the hypotheses is based on logical reasoning. 
This second step is similar to the scoring or ranking function used in the second step of an information retrieval process.

There are several ways to limit the number of potential hypotheses, for example, limiting the context, constraining the search space, or using a heuristic search method. Generally, only commonsense heuristic reasoning can help to limit the number of potential explanations. For example, suppose we want to explain why a health professional regulator received a complaint. In that case, we are only interested in explanations that are related to the health and care professionals' conduct or practice according to the relevant regulations \cite{lertvittayakumjorn2021supporting, jago2021use}. Therefore, to be useful in application, we need a mechanism to reduce the set of possible explanations.

Bhagavatula et al. \cite{bhagavatula2020abductive} introduce the \art{} dataset, which contains 20K commonsense narrative contexts and 200K explanations.  They approach abductive reasoning as the problem of finding the hidden middle of a linear series of the form, $\mathcal{O}_1 \land \mathcal{H} \implies \mathcal{O}_2 $, where we observe $\mathcal{O_2}$, and $\mathcal{O_1}$, and we try to come up with the best hypothesis $\mathcal{H}$. In one example, they show two observations:

\begin{itemize}
    \item \textbf{Observation 1:} Jenny cleaned her house and went to work, leaving the window just a crack open.
    \item \textbf{Observation 2:} When Jenny returned home she saw that her house was a mess!
\end{itemize}
In ${\alpha}NLI$, the task of a model is to choose the most plausible explanatory hypothesis among the given candidates. For example:
\begin{itemize}
    \item \textbf{Hypothesis 1:} A thief broke into the house by pulling open the window.
    \item \textbf{Hypothesis 2:} At work, she opened her window and the wind blew her papers everywhere.
\end{itemize}

Of the hypotheses above, if their common sense semantics are accepted, Hypothesis 1 is the most plausible for the two given observations.
However, if we consider the possible existence of enthymemes, the outcome may be different. For example, if ``the house'' in Hypothesis 1 does not refer to Jenny's house, then Hypothesis 1 becomes completely inappropriate as an explanation for the observations.
The existence of enthymemes affects the reliability of \art{} and ${\alpha}NLI$ for two reasons:
\begin{enumerate}
    \item Dataset \art{} is constructed by crowd-sourcing with the fact that different people will have different sets of implicit arguments to support their different decisions. 
    \item  As a consequence, in ${\alpha}NLI$ task, the model simply chooses a hypothesis which is more plausible than the other.
    That is, the model always outputs a \textbf{Yes} label and a \textbf{No} label for a pair of inputs, it does not happen that both hypotheses are plausible (or both implausible). In addition, within this setting, annotators may introduce their bias into the training and evaluation of the model.
\end{enumerate}

\section{Dataset Construction}
\label{sec:data}
While crowdsourcing can be an effective way to build a large dataset for evaluating the performance of deep learning models on tasks which require some logical capability \cite{bhagavatula2020abductive}, such as abductive reasoning, it is important to carefully verify the soundness and quality of the data. This can help to ensure that the dataset accurately represents the task and does not contain errors or biases that could impact the model's performance.

One way to increase the number of data points and ensure the soundness of the data is to use logic-based data generation techniques, such as symbolic reasoning or logical theorem proving. This can help to generate a larger number of high-quality examples of abductive reasoning that are consistent with the intended task and application domain.

Additionally, it is important to define the task clearly and accurately based on the characteristics of the data. If the task is not well-defined or does not align with the characteristics of the data, the model's performance may be difficult to interpret or may be difficult to explain. Carefully defining the task and choosing a suitable evaluation metric can help to ensure that the model's performance is accurately assessed and interpreted.

First, we precisely formulate the problem. With two observations $\mathcal{O}_1$, $\mathcal{O}_2$, and a hypothesis $\mathcal{H}$, the model needs to verify the validity of the Expression \ref{eq:abductive}.
\begin{equation}
\label{eq:abductive}
\mathcal{O}_1 \land \mathcal{H} \implies \mathcal{O}_2 
\end{equation}
We then analyze the two drawbacks of \art{} and subsequently construct \lart{} as an expanded and improved version based on negation rules and theory generators.

\subsection{Logical Consistency}
Compared with conventional programming languages, natural language have higher semantic tolerance, but therefore lower logical consistency. This is why, until recently, direct natural language-to-software conversion tools have relatively limited use. This well-known challenge of natural language use can also be an issue affecting the quality of crowdsourced datasets like \art{} and the models trained on it.
Sharing the same view on this issue,  \cite{Clark_2020} train their Transformer model with data generated from a logic-based program called a theorem generator. Expanding on this result, 
Gaskell et al. \cite{Gaskell_2022} introduce an adversarial framework to improve the logical consistency of these ``soft'' theorem provers. This is done by training a discriminator, which is then used to detect incorrect outputs from the theorem prover.
The crucial insight of this work is that a model trained with improved logical consistency can be applied to the task of soft theorem-proving with higher accuracy.

Without the support of a logic-based program in the data generation process, \art{} does not provide any guarantee of logical consistency of its content. For example, there are some samples in the dataset whose plausibility determination is based heavily on the enthymeme in the evaluator's knowledge base, which has the potential to introduce inconsistency.
For example, here is a sample from \art{}:
\begin{itemize}
    \item \textbf{Observation 1:} Ron started his new job as a landscaper today.
    \item \textbf{Observation 2:} Ron is immediately fired for insubordination.
    \item \textbf{Hypothesis 1:} Ron ignores his boss's orders and called him an idiot.
    \item \textbf{Hypothesis 2:} Ron's boss called him an idiot.
\end{itemize}
In other words, without a logic-based program in place, the data generated by \art{} may not be completely consistent or accurate. Additionally, the evaluator's own biases and knowledge can influence the plausibility of certain samples, leading to inconsistencies in the dataset. The example above, where Ron is fired for insubordination, produces two different hypotheses for why that might be the case, which further illustrates this potential for inconsistency in the data generated by \art{}.

To overcome this drawback, instead of only requiring the model to choose between two given candidate hypotheses, we reformatted the dataset and forced the powerful pretrained models to {\em predict} the binary label without limiting the number of candidates.
The negative samples are derived from the positive ones by using logical negation, which guarantees the positive triples contain the best hypotheses explaining the given observations.
This is an important distinction from the way the dataset was used in the original setting.
This adjustment can provide the ability to achieve logical consistency, and determine whether or not a candidate can be a valid hypothesis with the two given observations. 

\subsection{Observation-Hypothesis Interchangeability}
From $\mathcal{O}_1 \land \mathcal{H} \implies \mathcal{O}_2 $, we can deduce $\mathcal{H}  \land \mathcal{O}_1 \implies \mathcal{O}_2 $. 
In other words, the first observation and the hypothesis are interchangeable. More specifically, they are two events producing the second observation. Of the two events becomes the observation, while the event {\em we do not observe} becomes the hypothesis. However, in terms of logic, the two events hold equal footing, as they are both postulated to explain the second observation.
From this, in terms of dataset construction, we can double 
the number of positive samples by reversing the role of the hypothesis and the first observation. 

This logical reformulation helps us to realize that if we can not interchange the hypothesis and the first observation in a triple, the triple is not a valid abductive reasoning sample. For example, assume that we have a silly triple of $\mathcal{O}_1$, $\mathcal{H}$ and $\mathcal{O}_2$ as follows:  
\begin{itemize}
    \item  $\mathcal{O}_1$: John is the smartest person in the class.
    \item  $\mathcal{H}$: Every smart person has a green car.
    \item  $\mathcal{O}_2$: John has a green car.
\end{itemize}

In this case, we cannot interchange the $\mathcal{H}$ and $\mathcal{O}_1$ to get:
\begin{itemize}
    \item  $\mathcal{O}_1$: Every smart person has a green car.
    \item  $\mathcal{H}$: John is the smartest person in the class.
    \item  $\mathcal{O}_2$: John has a green car.
\end{itemize}

This is because the inference chain requires one more piece of information (i.e., the argument for $\mathcal{O}_2$ from $\mathcal{O}_1$ and $\mathcal{H}$ is an enthymeme), namely: ``The smartest person in the class is a smart person,'' which is not included in the triple.
In the latter triple, the hypothesis ``John is the smartest person in the class." is not reasonable given only the two observations ``Every smart person has a green car." and ``John has a green~car."

\subsection{Logic-augmented Dataset}
In our dataset construction process, we consider the above two reformulation factors which are not considered in \art{}: (1) We use a logic-based theorem generator to ensure logical consistency in the data; (2) We use logical formulas to ensure the validity of the triples in terms of abductive reasoning.
Based on these two transformations, we expand and improve the \art{} dataset and propose a new dataset called the \emph{logic-augmented abductive reasoning dataset (\lart{})}.
\lart{} is introduced as a dataset for a binary classification problem, where the model is trained to predict the validity of each provided triple.

As described in Section \ref{sec:ar}, in the \art{} dataset there are triples with high plausibility and others with low plausibility.  We select the highly plausible
triples as positive samples.
With the logic-based theorem generator, we randomly generate positive  samples
that are logically consistent and find the inference chains which have at least two inference steps to extract $\mathcal{O}_1$, $\mathcal{H}$ and $\mathcal{O}_2$. 
The hypotheses are then reversed to double the number of positive samples.

Producing negative samples in the context of abductive reasoning is more challenging than that in ordinary negation. We are looking for a hypothesis to explain the given observation; but we can not apply a random strategy to generate negative samples as there is no way for us to know whether a random hypothesis is reasonable for the given observations. 
We therefore limit the possibilities through our strategy of exploiting negation, but applying this strategy is not straightforward in the context of abductive reasoning. Consider that randomly negating the operators in the Expression 
\ref{eq:abductive} might still yield a triple labeled as true. We handle this issue by first constructing a truth table for the Expression \ref{eq:abductive}, and then uniformly negating the operators. The truth table for the Expression \ref{eq:abductive} is as in Table \ref{tab:truth_value}. 

\begin{table}
\centering
\caption{Truth table for Expression~\ref{eq:abductive}:  $\mathcal{O}_1 \land \mathcal{H} \implies \mathcal{O}_2 $. The first row corresponds to the truth values in the case of positive samples. } 
\begin{tabular}{|c|c|c|c|}

\hline
$\mathcal{O}_1$ & $\mathcal{H}$ & $\mathcal{O}_2$ &$\mathcal{O}_1 \land \mathcal{H} \implies \mathcal{O}_2 $ \\ \hline
\textbf{T}           & \textbf{T}          & \textbf{T}           & \textbf{T}   \\ \hline
F           & F          & F           & T                        \\ \hline
F           & F          & T           & T                        \\ \hline
T           & F          & F           & T                        \\ \hline
T           & F          & T           & T                        \\ \hline
F           & T          & F           & T                        \\ \hline
F           & T          & T           & T                        \\ \hline
T           & T          & F           & F                        \\ \hline
                     
\end{tabular}
\label{tab:truth_value}
\end{table}

The first row of Table \ref{tab:truth_value} corresponds to the truth values in the case of positive samples. We can easily see that the only logical option to negate Expression \ref{eq:abductive} is to negate the operator $\mathcal{O}_2$. Interestingly, when we interchange  $\mathcal{O}_1$ and  $\mathcal{H}$, the soundness of the system is not affected.
 This is because the soundness of the system is based on the logical equivalence of $ \mathcal{O}_1 \land \mathcal{H} \implies \mathcal{O}_2$ and  $ \mathcal{H} \land \mathcal{O}_1 \implies \mathcal{O}_2$. 

Compared to \art{}, the data format,  number of samples, and their logic consistency in \lart{} are significantly improved.
\art{}, for the positive (plausible) hypotheses, presents a narrative context to Amazon Mechanical Turk workers who were then asked to make assumptions and write natural language hypotheses for the two given observations. For the negative (implausible) hypotheses, the plausible hypothesis can be modified through minimal edits (up to 5 words).
For the positive (plausible) hypotheses, we use a logic-based theorem generator to randomly generate positive samples that are logically consistent, and identify the inference chains which have at least two inference steps to extract $\mathcal{O}_1$, $\mathcal{H}$ and $\mathcal{O}_2$. We also reuse the positive samples from \art{} and reverse the role of the hypothesis and the first observation. For the negative (implausible) hypotheses, we use a truth table to construct negative samples. The truth table shows that to negate the expression $\mathcal{O}_1 \land \mathcal{H} \implies \mathcal{O}_2 $, we need to change the value of $\mathcal{O}_2$ to false.
The model is trained to predict the validity of each provided triple.
The \lart{} dataset has almost 2.5 times as many samples as the \art{} dataset, which contains 200k samples. So 476,167 of the samples in \lart{} are used for training, 9,339 for validation, and 13,191 for testing.
This dataset can be used as a benchmark for measuring the abductive reasoning skills of state-of-the-art models in the legal domain, but it's not limited to this purpose, as it can also be a valuable resource for other natural language processing and machine learning tasks, especially to consider complex NLP tasks because of transformers lack of reasoning ability.
\section{Task Redefinition}
\label{sec:task_refine}
In addition to the \lart{} data's quantity and quality, task definition is also important in training and evaluating the model. 
Bhagavatula et al. \cite{bhagavatula2020abductive} introduce the ${\alpha}NLI$ task, in a way that the model needs to select the most plausible explanatory hypothesis between the two given. We argue that although this task is appropriate for evaluating the ability of the model to perform abductive reasoning and find the most plausible explanation, the way the authors limit the  ${\alpha}NLI$ task to a binary classification of two hypotheses makes the problem easier and can lead to an overfitting of the model. The model needs only to learn a binary classification of the two hypotheses and does not need to learn to find the most plausible explanation. 
They define ${\alpha}NLI$ task as follows:

\begin{itemize}
\item  $\mathcal{O}_1$ and $\mathcal{O}_2$ are two observations at time $t_1 < t_2$.
\item  $h+$  is a positive (plausible) hypothesis and $h-$  is a negative (implausible) hypothesis.
\item  The ${\alpha}NLI$ 
task is to select the most plausible hypothesis from the $h+$ and $h-$.
\end{itemize}
\textbf{}

We redefine  ${\alpha}NLI$ as ${\alpha}NLI^*$:

\begin{itemize}
    \item  $\mathcal{O}_1$ and $\mathcal{O}_2$ are two observations at time $t_1 < t_2$.
\item  $h$  is a candidate hypothesis.
\item The ${\alpha}NLI^*$ task is to test whether triple $\mathcal{(O}_1$, $h$, $\mathcal{O}_2$) is valid.
\end{itemize}

The approach of ${\alpha}NLI^*$ is similar to \textbf{${\alpha}NLI$} but different in the following ways:
\begin{itemize}
\item We ask the model to validate the triple instead of only choosing which hypothesis is more plausible amongst the given two;
\item In ${\alpha}NLI$, if the model did not choose a hypothesis, we still do not know whether it is valid, so we can only know that it is not plausible as the chosen one;
\item In addition, ${\alpha}NLI^*$ can be feasible even when only one hypothesis or more hypotheses are given, which is not possible in ${\alpha}NLI$.
\end{itemize}

\section{Experiments and Discussions}
\label{sec:experiment}
Our experiments are designed to test the performance of several alternative models for binary classification tasks related to abductive reasoning using the \lart{} dataset. Our extended dataset consists of 498,697 samples, of which 476,167 are used as the training set, 9,339 are used as the validation set, and 13,191 are used as the test set. The max-length in characters for observation 1, observation 2, and hypothesis in the training, validation, and test sets are shown in Table \ref{tab:maxlength}. Pre-trained transformer models have a built-in maximum token length for input, and ensuring that the maximum length does not exceed this limit helps avoid over-truncation issues. 
The number of samples in each class is identical in our binary classification setting.

\begin{table}[ht]
\centering
\caption{Max-length in characters for observation 1, observation 2, and hypothesis in the train, validation, and test sets.}
\begin{tabular}{lrrr}
\hline
             & Train & Validation & Test \\ \hline
Observation 1 ($\mathcal{O}_1$) & 71   & 71        & 71  \\
Observation 2 ($\mathcal{O}_2$) & 74   & 71        & 71  \\
Hypothesis    ($\mathcal{H}$) & 184   & 147        & 150  \\ \hline
\end{tabular}

\label{tab:maxlength}
\end{table}

In this experiment, it is crucial to use different training, validation, and test sets to ensure the evaluation of the selected transformer models is reliable and not influenced by overfitting to the training data. The training set is utilized for training the models, while the validation set is employed for tuning the hyperparameters and selecting the best model. Lastly, the test set assesses the final performance of the chosen model. 

We chose specific train/test/validation split ratios to ensure that the models have adequate data for training while still maintaining ample samples for validation and testing. It is worth noting that deep learning approaches like these are inherently statistical in nature, and our validation/test sets, with approximately 10K samples each, provide a reliable reflection of the results. 

We use several different transformer models in the experiment: the original BERT~\cite{devlin2018bert} (base and large version), and the top legal models (BERT-PLI~\cite{ijcai2020p484}, Legal BERT~\cite{chalkidis2020legal}, BERTLaw\cite{nguyen2020jnlp} and NFSP version of Paralaw NetsParaLaw Nets~\cite{nguyen2021jnlp}).
We train each model on the input data, using the valid and invalid triples as the labels, and evaluate the performance of the models on the test set using accuracy. We also report the performance on our validation set. We run the experiment multiple times to ensure the reliability of the results, and we analyze the performance of the models on the test data to determine which model performs best on the binary classification tasks related to abductive reasoning. 
We also tested GPT-3 \cite{brown2020language} zero-shot prediction and recorded the results on the test set of \lart{}.

\begin{table}[ht]
\caption{Average performance of the models in three runs, sorted from high to low.}
\label{tab:results}
\begin{center}
\begin{tabular}{lcr}
\hline
Model & Validation Accuracy & Test Accuracy \\
\hline
BERT Base & 0.6202 & 0.6162 \\
BERT-PLI & 0.6115 & 0.6115 \\
NFSP & 0.5825 & 0.5808 \\
Legal BERT & 0.5756 & 0.5619 \\
BERTLaw & 0.5484 &  0.5371 \\
BERT Large &0.5016&0.5000 \\
GPT-3 &-&0.4959\\
\hline
\end{tabular}
\end{center}
\end{table}

Table \ref{tab:results} shows the results of the experiment. 
Our first observation is that performance on the validation set and the test set are not significantly different, which is a good indication that the models aren't overfitting the training data or the hyperparameter tuning process. From those results, we can also observe that the original BERT Base model performs better on abductive reasoning than all state-of-the-art legal models and even the BERT Large model. This result is quite surprising because it means that the pre-trained legal models are not necessarily more effective than the original BERT Base model on abductive reasoning.

We believe this is because the legal models are trained on documents that are mostly directed toward legal reasoning, rather than abductive reasoning. In the legal domain, there are many documents that contain information that can help with legal reasoning, but there is not a lot of information in the legal domain that can help with abduction. This imbalance in the training data can lead to a bias in the training process that favors legal reasoning and harms the performance of the models on abductive reasoning. 


The worst performance among finetuned models is BERT Large, which is also a surprise since it is usually reported to be a robust model on many NLP tasks. 
This result suggests that pretraining a model with a larger capacity with more data does not guarantee better performance. 
Adding commentary on GPT-3, which has the lowest performance on zero-shot tasks, we included it for reference purposes, not for comparison, as the model has not been tuned for the specific domain and therefore cannot be fairly compared to the other models. Furthermore, a general comment can be made that the poor performance of these models indicates that abductive reasoning remains a challenging problem. The redefinition of the task within \lart{} helps to make this issue more evident.

\section{Conclusions}
We have investigated the support for abductive reasoning provided by state-of-the-art (SOTA) transformer models within the legal field. To accomplish this, we first redefined the task of abductive reasoning and constructed a reliable dataset. Following this, we utilized the dataset to assess the performance of SOTA models in the legal sphere, as well as prominent large language models in Natural Language Processing (NLP).
Our experimental results revealed that the SOTA models, including all legal-specific variants, do not necessarily outperform the original BERT Base model in abductive reasoning tasks. This outcome provides insight into current limitations when pretraining large language models for legal applications.
The subpar performance of the original BERT Large and GPT models illustrates that simply increasing the size of the model and providing it with more data does not guarantee superior performance.
Future directions for this research could include exploring alternative pretraining approaches specifically tailored to abductive reasoning tasks, developing novel architectures that focus on legal reasoning, and examining the relationship between model capacity and performance on abductive reasoning tasks. Additionally, further investigation into leveraging and integrating existing legal domain knowledge with the pretraining process may lead to more effective models capable of handling the unique challenges of legal reasoning tasks. 


\begin{acknowledgments}
  This work was supported by JSPS KAKENHI Grant Number, JP22H00543 and JST, AIP Trilateral AI Research, Grant Number JPMJCR20G4.
  Francesca Toni and Kostas Stathis would like to thank the National Institute of Informatics, Tokyo, Japan, for supporting their visit to Japan that made this work possible. Francesca Toni also acknowledges support from the European Research Council (ERC) under the European Union’s Horizon 2020 research and innovation programme (grant agreement No.101020934, ADIX), as well as support from J.P. Morgan and the Royal Academy of Engineering, UK, under the Research Chairs and Senior Research Fellowships scheme. 
\end{acknowledgments}

\bibliography{sample-ceur}

\appendix



\end{document}